\documentstyle[12pt,epsf]{article}
\newlength{\abstractwidth}
\renewcommand{\thefootnote}{\fnsymbol{footnote}}
\renewcommand{\thanks}[1]{\footnote{#1}} 
\newcommand{\starttext}{
\setcounter{footnote}{0}
\renewcommand{\thefootnote}{\arabic{footnote}}}
\renewcommand{\theequation}{\thesection.\arabic{equation}}
\newcommand{\be}{\begin{equation}}
\newcommand{\bea}{\begin{eqnarray}}
\newcommand{\eea}{\end{eqnarray}}
\newcommand{\beq}{\begin{equation}}
\newcommand{\ee}{\end{equation}}
\newcommand{\eeq}{\end{equation}}

\newcommand{\I}{{\cal I^+}}
\newcommand{\It}{\tilde {\cal I}^+}
\newcommand{\la}{\lambda}
\newcommand{\pp}{\partial}

\def\[{\left [}
\def\]{\right]}
\def\({\left (}
\def\){\right)}

\begin{document}
\renewcommand{\theequation}{\thesection.\arabic{equation}}
\begin{titlepage}
\bigskip
\rightline{}  \rightline{NSF-ITP-01-38}
 \rightline{hep-th/0105111}
\bigskip\bigskip\bigskip\bigskip
\centerline{\Large \bf {Fate of the Black String Instability}}
\bigskip\bigskip
\bigskip\bigskip

\centerline{\large Gary T. Horowitz${}^{1,2}$ and  Kengo Maeda${}^1$}
\bigskip\bigskip
\centerline{\em ${}^1$Department of Physics, UCSB, Santa Barbara, CA. 93106}
\medskip
\centerline{\em ${}^2$Institute for Theoretical Physics, UCSB, Santa Barbara,
CA 93106}
\medskip
\bigskip\bigskip


\begin{abstract}
Gregory and Laflamme showed that certain
nonextremal black strings (and $p$-branes) are
unstable to linearized perturbations. It is widely believed that this
instability will cause the black string  horizon to
classically pinch off and then quantum mechanically separate, resulting in
higher dimensional black
holes. We argue that this cannot happen. Under very mild assumptions,
classical event horizons cannot pinch off. Instead, they settle down
to new static black string solutions which are not translationally
invariant along the string.

\medskip
\noindent
\end{abstract}
\end{titlepage}
\starttext \baselineskip=18pt \setcounter{footnote}{0}

\setcounter{equation}{0}
\section{Introduction}

It is well known that four dimensional black holes are stable \cite{chand}. 
Almost ten
years ago,
Gregory and Laflamme \cite{grla1,grla2}
showed that this is not true for higher dimensional
generalizations of black holes, such as black strings and black $p$-branes.
The simplest black string solution is
the product of the four dimensional Schwarzschild metric and a circle of length
$L$. Gregory and Laflamme showed that this spacetime is unstable
to linearized perturbations with a wavelength along the circle larger than
the Schwarzschild radius of the black hole $r_0$. They also compared
the total entropy of the black string with that of a five dimensional black hole
with the same total mass,
and found that when $L>r_0$, the black hole had greater entropy. They thus
suggested that the full nonlinear evolution of the instability would result
in the black string breaking up into separate black holes which would then
coalesce into a single black hole. Classically, horizons cannot bifurcate, but
the idea was that under classical evolution,
 the event horizon would pinch off and become
singular. When the curvature became large enough, it was plausible that
quantum effects would smooth out the transition between the black string
and black holes. The same instability was found for higher dimensional
black $p$-branes, and also black strings (and $p$-branes) 
carrying certain charges,
as long as they were  nonextremal.

These higher dimensional generalizations of black holes arise naturally
in string theory, and
the idea that a black string with $L> r_0$
will break up into black holes has been  widely accepted. It has been used
in many
recent string discussions, e.g.,
descriptions of black holes in matrix theory \cite{bfks,homa,suss},
discussions of the
density of states of strongly coupled field theories
using the AdS/CFT correspondence \cite{bdhm, lms, bkr}, the relation
between near extremal D2 and M2 brane configurations \cite{imsy}, 
discussions
of black holes on brane-worlds \cite{chr}, and various unstable D-brane
configurations \cite{hyak,hemg}.

We will argue that this widespread belief is
incorrect: Black strings do not in fact,
break up into black holes! Under very weak assumptions, we prove that
an event horizon cannot pinch off in finite time. In particular, if one perturbs
Schwarzschild cross a circle, an $S^2$ on the horizon cannot  shrink to
zero size in finite affine parameter. 
The basic idea is the following. The famous area theorem
is based on a local result that the divergence $\theta$ of the null geodesic
generators of the horizon cannot become negative. If an  $S^2$ on the horizon
tries to shrink to zero size, $\theta$ can stay positive only if the horizon
is expanding rapidly in the circle direction. But this produces a large shear
which also drives $\theta$ negative. The upshot is that 
the solution settles down to a new (as yet unknown)
static black string solution which is not translationally invariant along
the circle\footnote{There is a slight possibility that the horizon 
continues to pinch off, taking an
infinite time to do so. If this happened, the curvature would eventually become
large and  quantum effects could cause the horizon to split. 
We will argue that this possibility is very unlikely, but have not
been able to rigorously exclude it.}.

One can view this result as an example of spontaneous symmetry breaking
in general relativity.
The most symmetric solution is unstable, and the
stable solution has less symmetry.        
Unlike the usual particle physics examples
where the broken symmetry is an internal one, here
the broken symmetry is spatial translations. 

Our arguments apply to all black $p$-branes. If the $p$-brane is charged with
respect to a $p+2$ form $F$ (so the charge is the integral of $*F$ over a
sphere surrounding the brane), it was already known that the $p$-brane
could not break up into black holes since this charge can only be carried
by an object extended in $p$ directions. Since nonextremal solutions are
still unstable, it was clear that there must be new static solutions with
less symmetry. However, these were thought to resemble the extremal
solution with a Schwarzschild black hole superposed on it. Since the
extremal $p$-brane horizon is often singular, this cannot happen. There
must be new solutions with nonsingular horizons.
Our results also apply to black $p$-branes with ``smeared" charges
associated with a lower rank form. Charge conservation does not prevent these
solutions from breaking up, but nevertheless, the horizon must stay connected.

\section{Horizons cannot pinch off}

For simplicity, we will start with the example of four dimensional Schwarzschild
with radius $r_0$ cross a circle of length $L>r_0$. Since it has been
suggested that cosmic censorship might be violated in the evolution of the
Gregory-Laflamme instability (see below), we
must be careful not to assume cosmic censorship in our analysis. We will
proceed by considering the maximal Cauchy evolution of smooth initial data
on a surface $\Sigma$.
Consider initial data for a vacuum spacetime
which looks
like a static slice of the black string plus a small perturbation.
Alternatively, to avoid the second asymptotically flat region in the
maximal extension
of Schwarzschild, one can start with initial data describing infalling matter
that will produce trapped surfaces and an apparent horizon with $S^2 \times
S^1$ topology. Now consider the maximal Cauchy evolution of this initial data.
Since the initial data is asymptotically flat, the evolution will include
at least part of null infinity $\I$. Let $\It = \I \bigcap D^+[\Sigma]$
where $D^+$ denotes the future domain of dependence in the conformally 
completed spacetime.
 If the weak energy condition
is satisfied, the trapped surfaces cannot lie in the past of $\It$. Thus
there must be an event horizon, defined by the boundary of $I^-[\It]$,
enclosing all the trapped surfaces.
We want to study properties of this event horizon.

\begin{figure}[htbp]
 \centerline{\epsfxsize=10.0cm \epsfbox{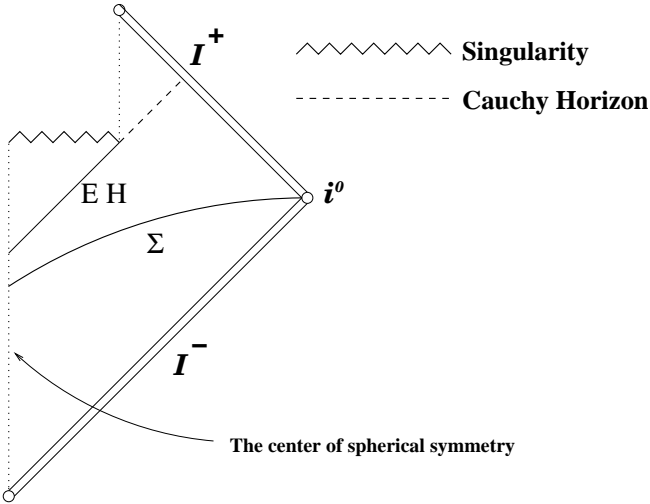}}
        \caption{\sl A Penrose diagram of $z=0$ surface if horizon pinches
         off. We show that this cannot happen.}
             \protect
\label{fig1-eps}
\end{figure}

To begin, we will consider the simplest case when the spacetime has spherical
symmetry, and also reflection symmetry about $z=0$, where $z$ is the coordinate
along the $S^1$. To picture the evolution, consider the metric restricted to
$z=0$. This looks like a time dependent
four dimensional spherical black hole. If the horizon shrinks to zero
size in finite affine parameter, 
the Penrose diagram would look like Fig.~\ref{fig1-eps}.

The Cauchy evolution
stops at the dashed line. The spacetime to the future of this line
assumes the horizon splits into separate black holes. Note that the
spacetime looks just like an evaporating black hole and has a
naked singularity. However we are now
considering classical evolution. The horizon size on the $z=0$ slice
can decrease because there is an effective stress energy tensor coming from
the higher dimensional curvature which can have negative energy densities.
However, we now show that it cannot shrink all the way to zero size in
finite affine parameter.

In a neighborhood of the horizon one can introduce Gaussian null coordinates
so that the metric takes the form \cite{frw}
\be\label{horizon}
ds^2 = - fd \la^2 + 2dr d\la + 2\beta  d\la dz +e^{2\chi} dz^2 + e^{2\psi} d\Omega
\ee
where $f,\beta,\chi,\psi$ are functions of $\la,r,z$. The horizon is located at
$r=0$ and on the horizon, $f=0, \  \pp_r f=0, \ \beta=0$ and $\la$
 is an affine parameter
along the null geodesic generators. Let $\ell^\mu = (\pp/\pp \la)^\mu$ 
be tangent to 
these
null generators, and let $h_{\mu\nu}$ denote the metric on a $r=0$, constant
$\la$ cross-section of the horizon\footnote{The indices $\mu,\nu$
run over all spacetime coordinates, but $h_{\mu\nu}$ is nonzero only 
when $\mu,\nu$ is $z$ or one of the angles on the spheres of spherical
symmetry.}.
The divergence of the null generators is
\be\label{thetadef}
\theta =  h^{\mu\nu}\nabla_\mu \ell_\nu = \dot \chi + 2\dot \psi
\ee
where a dot denotes derivative with respect to $\la$. 
A fundamental property of event horizons is that $\theta$ cannot become negative.
This can be proved under two different assumptions. The first is that the null
geodesic generators of the horizon are complete. We do not wish to assume this
since, if the horizon pinches off in finite time, the geodesics will
not be complete. However, $\theta \ge 0$ can also be established if there are no
naked singularities outside the horizon \cite{wald}. Since we are working
within the future domain of dependence of  a spacelike surface $\Sigma$,
the spacetime is globally hyperbolic and this assumption is valid. 

The shear of the null generators is defined by
\be
 \sigma_{\mu\nu} = {h_\mu}^\alpha {h_\nu}^\beta\nabla_\alpha \ell_\beta
-{\theta\over 3} h_{\mu\nu}
\ee
which reduces to 
\be
\sigma_{\mu\nu} = {1\over 2} \dot h_{\mu\nu} -
{\theta\over 3} h_{\mu\nu}
\ee
 For the metric (\ref{horizon}) one
finds
\be
\sigma_{\mu\nu} \sigma^{\mu\nu} = {2\over 3} (\dot \chi - \dot \psi)^2
\ee
Since $\theta \ge 0$ and $\dot\psi \le 0$, we have
 $\dot\chi \ge -2\dot \psi$ and $\sigma_{\mu\nu} 
\sigma^{\mu\nu} \ge 6 \dot
\psi ^2$. The Raychaudhuri equation in five dimensions is 
\be\label{raych}
\dot \theta = -\theta^2/3 - \sigma_{\mu\nu} \sigma^{\mu\nu} -
 R_{\mu\nu} \ell^\mu \ell^\nu
\ee
Since the spacetime is Ricci flat, we have $\dot \theta \le -6\dot \psi^2$. Thus
if $\theta_0>0$ is the initial value of the divergence,
\be
\theta(\la) \le \theta_0 -6\int_0^\la \dot \psi^2
\ee
Using $(\dot \psi +1)^2 \ge 0$, this implies $\theta(\la) \le 12 \psi (\la) 
+6\la +$
constant.
Since $\theta$ must stay positive,  $\psi$ cannot go to minus
infinity at any finite $\la$. In other words, the
sphere cannot shrink to zero size in finite affine parameter.

The above argument can be extended in many directions. One can consider
nonspherical perturbations, higher dimensional spacetimes, horizons 
extended in more than one direction (black branes), and collapsing
surfaces of various dimensions. Perhaps the most general result is the
following theorem which states roughly that event horizons cannot have
any collapsing $S^1$'s.

{\it Theorem}: Consider the event horizon of a general $D$ dimensional
spacetime satisfying the null energy condition. Choose any $S^1$
in the $\la=0$ cross-section of the horizon. Following this circle along the
null generators yields a family of
closed curves $\Gamma(\la)$. 
The length of these curves cannot go to zero in 
finite affine parameter.

{\it Proof}: 
Choose local coordinates $y,x^i$ ($i = 1, \cdots, D-3$) on the horizon
in a neighborhood of $\Gamma(\la)$ which are
constant along the null generators, such that the $\Gamma(\la)$ is $x^i=0$,
and $y$ is a periodic coordinate along the curves.
The metric $h_{\mu\nu}$ on each constant $\la$ cross-section of the horizon
can be written
\be\label{hormet}
ds^2 = e^{2\phi}(dy + A_i dx^i)^2 + \gamma_{ij} dx^i dx^j
\ee
where $\phi,A_i$ and $\gamma_{ij}$ depend on  $\la$ as well as the other
coordinates.
Let $\ell^\mu$ be the null geodesic generators of the event horizon and
set $B_{\mu\nu} = {h_\mu}^\alpha {h_\nu}^\beta \nabla_\alpha \ell_\beta$.
Then the Raychaudhuri equation, in any dimension, is simply
\be\label{genrach}
\dot \theta = -B_{\mu\nu}B^{\mu\nu} - R_{\mu\nu} \ell^\mu\ell^\nu
\ee
Decomposing $B_{\mu\nu}$ into its trace $\theta$ and tracefree $\sigma_{\mu\nu}$
parts we recover (\ref{raych}) when $D=5$. 
Using (\ref{hormet}), a straightforward calculation yields
\be
B_{\mu\nu}B^{\mu\nu} = \dot\phi^2 + {1\over 2} e^{2\phi}||\dot A_i||^2
+{1\over 4}||\dot \gamma_{ij}||^2 
\ee
where the double bar  means take the norm with the metric $\gamma_{ij}$.
Since the weak energy condition holds, (\ref{genrach}) implies
\be
\dot \theta \le -\dot\phi^2
\ee
The previous argument now shows 
 that $\phi $ cannot go to minus infinity in  finite
$\la$. This completes the proof.

An immediate consequence of this theorem  is that
cosmic censorship is not violated by the Gregory-Laflamme instability.
The null geodesic generators of the horizon remain complete. 

\setcounter{equation}{0}
\section{Discussion}

Since the horizon cannot pinch off in finite affine parameter
the spacetime
 must settle down to something at late time. The most likely possibility
is that it settles down to a new static black string (or $p$-brane)
solution which is not translationally invariant along the horizon.
Before discussing this in more detail, we comment on the remote
possibility that the horizon pinches off in infinite affine parameter.
Using just the Raychaudhuri equation, it is possible for the length 
${\cal L}(\la)$
of the closed curves $\Gamma(\la)$ to go to zero keeping $\theta$ positive.
However, since the horizon area must remain finite, $\int^\infty \theta(\la)d\la
<\infty$, and the decay rate is
quite restricted. For example, it is easy to see that ${\cal L}(\la)$
cannot decay like a simple power law or exponential. The type of decay that
is not obviously forbidden is 
\be
{\cal L}(\la) = e^{-(\ln \la)^\alpha} \qquad \qquad 0<\alpha < 1/2
\ee
This seems rather unnatural. More importantly, it is 
 a very slow decay.  
Since the decay is so slow, there must exist 
a family of new essentially static solutions. One can view the late time
evolution
as slow motion through this space of static solutions. But given
the existence of new static solutions, there is no physical reason 
for the horizon
to pinch off. It is
much more likely that the evolution will stop at one of the static 
configurations.

In the case of Schwarzschild cross a circle, one can say  more. Consider
the metric (\ref{horizon}) and assume reflection symmetry about 
$z=0$. If the horizon pinches off in infinite
time then $\psi(z=0,\la) \rightarrow -\infty$ as
$\la\rightarrow \infty$. Since $\theta\ge 0$ everywhere, (\ref{thetadef})
implies that $\chi$ must
go to plus infinity. Thus it appears that the solution near $z=0$ is
evolving toward a very long thin black string, which is unstable. This
not only sounds unphysical, it leads to a contradiction by looking at
the effective four dimensional Einstein's equation on the $z=0$ surface.
By using a Kaluza-Klein type reduction in the $z$ direction (which involves
rescaling the effective 4D metric),
 the equations take the form of
Einstein gravity minimally 
coupled to a massless scalar $\chi$. Spherically symmetric black hole
solutions always have $\chi$ remain bounded on the horizon. 
This contradicts the fact
that $\chi$ must diverge.
Unfortunately this is not 
sufficient to rule out the possibility that the horizon
pinches off in infinite time. If $\psi_{,zz}|_{z=0}$ 
grows sufficiently rapidly with
$\la$, then the solution never resembles a long thin string. Instead
it looks like a chain of spherical black holes connected by small necks. 
But the spacetime near one neck would be analogous to that obtained by
 bringing two black holes close 
together, and in that case it is well known that a new  trapped surface forms
which surrounds both black holes\footnote{We thank V. Hubeny for
suggesting this analogy.}. This is simply because there is now
double the mass within a sphere containing both black holes so the effective
Schwarzschild radius moves out. Similarly, we would expect that
if the apparent horizon tried to pinch off in infinite affine parameter, 
another apparent horizon would form outside, and the true event horizon
would not pinch off.

It should be noted that the argument about Kaluza-Klein reduction in the $z$
direction also rules out the possibility that the final static solution
is $z$ independent, but has $\chi(r)$ increasing with $r$ such that
the length of the circle at the horizon
is less than $r_0$. Since the 5D Einstein's equation reduces to 4D gravity
coupled to a  scalar $\chi$, the usual no hair theorem shows that a static
black hole must have constant $\chi$.

Let us now turn to the most likely alternative that the solution
settles down to a
new static black string solution of Einstein's equation which is 
not translationally invariant along the horizon. Since we do not have
the new solution explicitly, we cannot say for sure what the horizon
geometry looks like. In particular, we cannot determine the minimum size of
the sphere. 
 However, since the
physics is mostly determined by one scale\footnote{The length of the
circle $L$ is not likely to be important when $L \gg r_0$ since the
transition between stable and unstable modes is set by $r_0$.}
 -- the initial Schwarzschild radius $r_0$ --
the maximum and minimum radii are likely to be within factors of two
of this scale. There is no reason for a large dimensionless number
(corresponding to the ratio of these radii) to 
arise. 
Note that one cannot determine the geometry of the horizon 
just by examining Einstein's equation near the
horizon. Even for ordinary four dimensional black holes, if there is
a static nonspherical distribution of matter far outside the black hole
the horizon will be distorted. One needs to examine the field equations 
everywhere.

Even without the exact solutions, one can deduce certain properties 
of the new solutions. For example,
the solution must approach the translationally invariant
black string exponentially fast at infinity. This is because the asymptotic
solution can be modeled by a perturbation of the translationally
invariant one. Since $z$ is periodic,
any $z$-dependent perturbation
satisfies a massive spin two  equation and must fall off
exponentially.

The translationally invariant black string is unstable for $L> L_0$
where $L_0$ is a critical length of order $r_0$.
For every initial Schwarzschild radius $r_0$ and $L>L_0$, there
must be at least one new static black string solution. There may even 
be more than one.
Since the black strings with $L<L_0$ are stable, the new black string
solutions probably  meet the
branch of solutions corresponding to the original translationally invariant
solutions at $L=L_0$. This is supported by the existence of a nontrivial static
 perturbation
to the translationally invariant black string when $L=L_0$ \cite{grla1}.

If $\xi$ denotes the static Killing field, then ${}^* d\xi$ is closed by
the vacuum Einstein equation. Consider the integral of the curl of this 
$(D-2)$-form over
a static slice from the horizon to infinity.
The surface term at infinity yields the total mass $M$, and the
surface term at the horizon yields $\kappa A/4\pi$ where $\kappa$ denotes
the surface gravity. Since these must be equal we obtain \cite{wald}
\be\label{mka}
M = {\kappa A\over 4\pi}
\ee
This is true for any static vacuum solution with a horizon.
If one starts with a slightly perturbed translationally invariant black string,
under evolution the mass decreases (since energy can be radiated to infinity)
and the area increases, so the final surface gravity cannot be greater than
the initial surface gravity. For nonvacuum spacetimes corresponding to 
charged branes, there is an extra term on the right hand side of (\ref{mka})
involving a volume integral of the stress energy tensor.

The final black string solution will have greater entropy than the original one
but considerably less than a single higher dimensional black hole (when
$L\gg r_0$). This can be seen by the following rough argument. 
Setting $G_5=1$, the original black string
has mass of order $L r_0$ and entropy of order $L r_0^2$. Since the horizon
cannot pinch off, the final black string must have an entropy less than
that of a chain of five dimensional black holes of radius approximately $r_0$.
Since each one has mass of order $r_0^2$ (and the total mass is $L r_0$),
 there are $L/r_0$ black holes
in the chain with total entropy of order $(L/r_0) r_0^3 = L r_0^2$.
 So the entropy
can only increase by a numerical factor. This is much less than the entropy
of a single five dimensional black hole with the same mass which would
be $(L r_0)^{3/2}$. However, 
this does not contradict the statement that the new solutions are stable.
There are many examples of stable solutions which do not maximize
their entropy for given mass, e.g. static stars, tables, chairs, etc. 
The new black string solutions can be viewed as a  local entropy maximum 
but not a global one. 
It has recently been suggested that a 
Gregory-Laflamme type instability should exist precisely when the entropy
is not a local maximum \cite{gumi,real}.

Clearly it would be very interesting to find the new black string 
solutions explicitly. Although an analytic solution would be preferable,
numerical solutions may be needed. Explicit numerical evolution of the
Gregory-Laflamme instability is now underway \cite{chop}.

The implications of these new solutions for string theory
 remain to be investigated. Previous results which assumed
black strings will break up into black holes should be reexamined. Some of
these
results will be unchanged, e.g., there is independent evidence for localized
black holes bound to branes \cite{ehm,gkr} so the assumptions of \cite{chr}
are not needed\footnote{There are also stable black strings with asymptotically
AdS boundary conditions \cite{hika}.}. Other results may need to be modified.

\vskip .5cm \centerline{\bf Acknowledgments} \vskip .5cm

It is a pleasure to thank V. Hubeny, N. Itzhaki, W. Unruh and R. Wald
 for discussions.
G.H. is supported in part by NSF grants
PHY-0070895 and PHY-9907949.
K.M. is supported by a JSPS Postdoctoral Fellowship for Research Abroad.

\end{document}